\documentclass[sigconf,10pt,letterpaper,twoside]{acmart}

\renewcommand\footnotetextcopyrightpermission[1]{} % remove copyright space
\title{ANSC: Probabilistic Capacity Health Scoring for Datacenter-Scale Reliability}

\author{Madhava Gaikwad}
\affiliation{%
  \institution{Microsoft}
  \city{Redmond}
  \state{WA}
  \country{USA}
}
\email{mgaikwad@microsoft.com}

\author{Abhishek Gandhi}
\affiliation{%
  \institution{Microsoft}
  \city{Redmond}
  \state{WA}
  \country{USA}
}
\email{abgandhi@microsoft.com}

\begin{document}

\begin{abstract}
We present ANSC, a probabilistic capacity health scoring framework for hyperscale datacenter fabrics. While existing alerting systems detect individual device or link failures, they do not capture the aggregate risk of cascading capacity shortfalls. ANSC provides a color-coded scoring system that indicates the urgency of issues \emph{not solely by current impact, but by the probability of imminent capacity violations}. Our system accounts for both current residual capacity and the probability of additional failures, normalized at datacenter and regional level. We demonstrate that ANSC enables operators to prioritize remediation across more than 400 datacenters and 60 regions, reducing noise and aligning SRE focus on the most critical risks.
%\footnote{All experiments are conducted on infrastructure telemetry and simulated incidents; no human subjects or sensitive data are involved, and therefore no ethical issues arise.}
\end{abstract}

\maketitle

\section{Introduction}
Hyperscale datacenters operate thousands of Clos-networked devices and links. Current incident management pipelines rely on alerts for direct failures (e.g., device down, link CRC, bandwidth exhaustion). While effective for triggering immediate repairs, these alerts are \emph{binary} and often miss broader systemic risks: when cumulative capacity reduction across a layer or datacenter could soon breach safe operational thresholds.

We propose the \textbf{Aggregate Network Safety Color (ANSC)} score. ANSC provides a probabilistic and capacity-aware risk score for each datacenter and region, allowing operators to forecast potential impact and prioritize mitigation. Unlike existing alerting, ANSC is not a direct measure of customer impact, but a proactive signal of reliability posture.

\section{Background and Motivation}
\subsection{Clos Topology in Datacenters}
\begin{figure}[h]
  \vspace{-0.2cm}   % tighten space above figure
  \centering
  \includegraphics[width=0.68\linewidth]{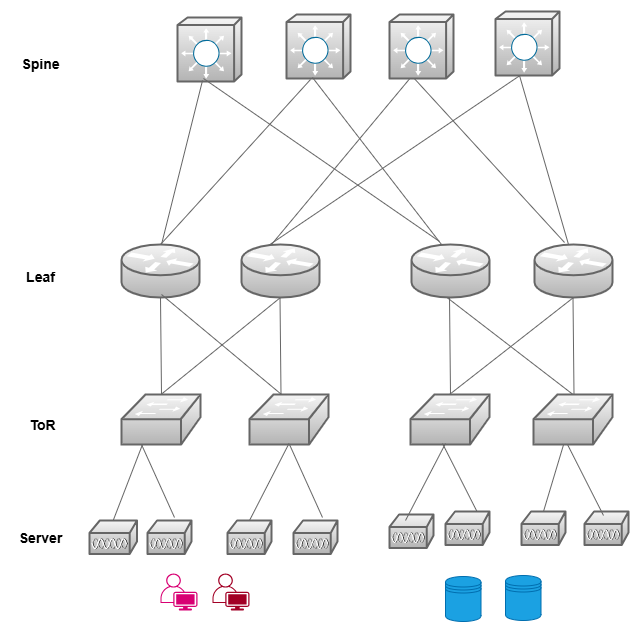}
  \vspace{-0.2cm}   % tighten space between image and caption
  \caption{Simplified Clos topology context for ANSC.}
  \vspace{-0.2cm}   % tighten space between image and caption
  \label{fig:clos}
\end{figure}

Modern datacenters employ Clos networks with thousands of servers, top-of-rack switches, multiple aggregation layers, and spine routers. Figure~\ref{fig:clos} illustrates a simplified view.

\subsection{Limitations of Current Safety Policy Engines}
Current ``safety policy engines'' check whether a device or link can be removed from production without violating capacity headroom. If insufficient headroom exists, the incident is escalated to an SRE for triage. This ensures immediate protection but does not forecast longer-term systemic degradation. For example, repeated link losses across a layer can accumulate until residual capacity is critically low.

\section{System Design}
\subsection{ANSC Scoring Model}
ANSC assigns each datacenter or region a score, mapped to color-coded states (red, orange, amber). Importantly, ANSC is not a direct measure of current outage impact, but rather:
\begin{itemize}
    \item Current effective capacity relative to forecasted demand.
    \item Probability of further degradation, given historical failure rates.
    \item Normalization against yearly safety budgets per datacenter and per region.
\end{itemize}

\subsection{Formula Flow}
Let $C_{avail}$ denote currently available capacity in a Clos layer, and $C_{req}$ the forecasted demand. Let $P_{fail}$ represent the probability of additional device/link failures, estimated from historical incident rates. The effective safety margin is:

\[
ES = \frac{C_{avail} - C_{req}}{C_{req}}
\]

\begin{figure}[h]
  \vspace{-0.2cm}
  \centering
  \includegraphics[width=0.9\linewidth]{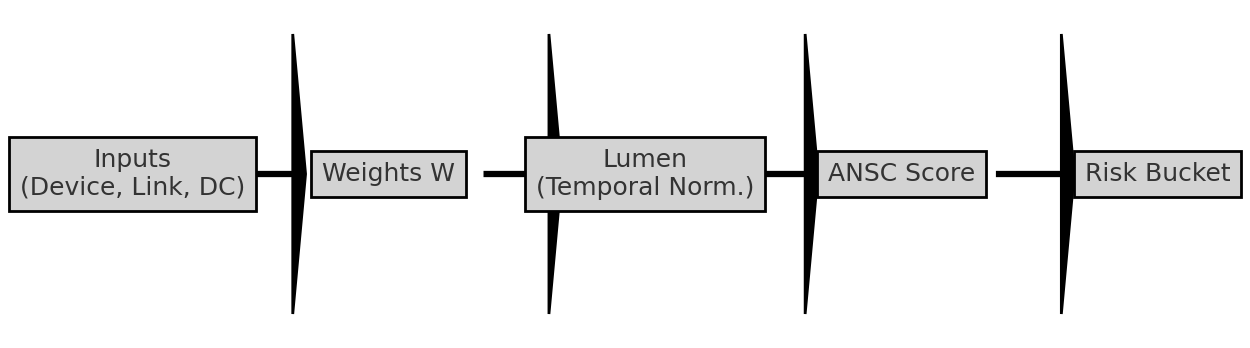}
  \vspace{-0.2cm}
  \caption{ANSC formula flow (simplified).}
  \vspace{-0.2cm}
  \label{fig:flow}
\end{figure}

The normalized ANSC score is then:

\[
ANSC = f(ES, P_{fail}, T_{pers})
\]

where $T_{pers}$ is the yearly persistence budget and $f(\cdot)$ maps to categorical color states. A schematic of this formula flow is shown in Figure~\ref{fig:flow}.

In practice, ANSC scores are not consumed in isolation. Operators routinely correlate them with existing telemetry dashboards (latency, error counters) to validate whether an elevated score reflects an emerging systemic risk. This cross-check provides confidence that ANSC complements rather than replaces existing monitoring, and helps build operator trust in acting on its recommendations.

\section{Methodology}
\subsection{Probability Estimation}
Failure probability $P_{fail}$ is estimated per link and device using historical outage rates, weighted by environmental and operational conditions (e.g., recent maintenance activity). At layer-level, we aggregate probabilities assuming partial independence.

\subsection{Normalization Across 400+ Datacenters}
To avoid over-sensitivity, we constrain the maximum fraction of red/orange/amber assignments annually: 5\%, 12\%, and 20\% respectively, with $\pm$5\% tolerance margin. This ensures only the most at-risk sites are escalated, while maintaining global prioritization fairness.

\section{Evaluation}
\begin{figure}[h]
  \vspace{-0.2cm}
  \centering
  \includegraphics[width=0.6\linewidth]{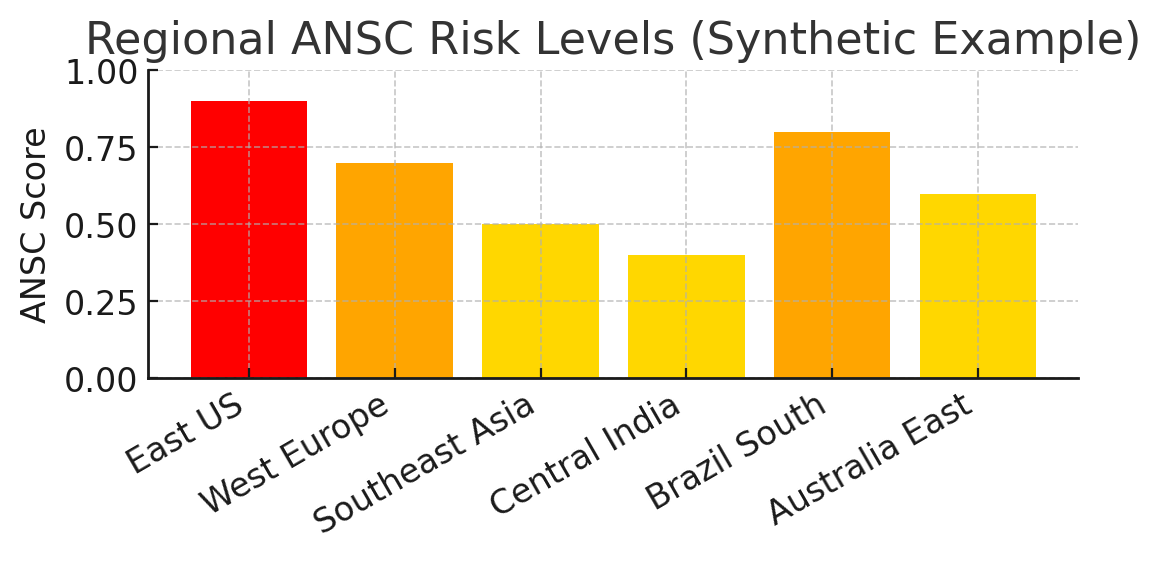}
  \vspace{-0.2cm}
  \caption{Regional ANSC heatmap example across Azure-like regions.}
  \vspace{-0.2cm}
  \label{fig:heatmap}
\end{figure}

We simulated ANSC scoring across 400 datacenters in 60 regions. Figure~\ref{fig:heatmap} shows an illustrative regional heatmap.

\begin{figure}[h]
  \centering
  \begin{minipage}{0.6\linewidth}
    \includegraphics[width=\linewidth]{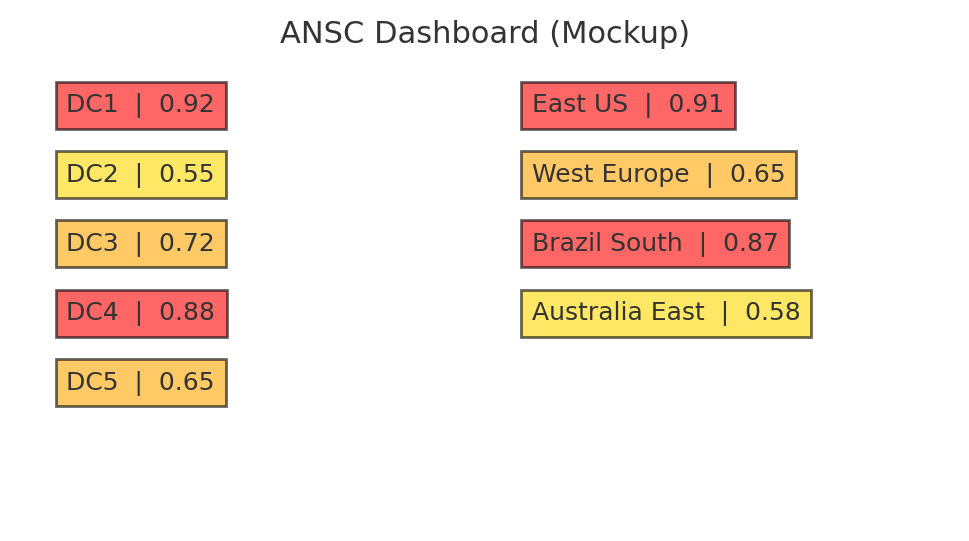}
  \end{minipage}\hfill
  \begin{minipage}{0.35\linewidth}
    \caption{ANSC operator dashboard mock.}
    \label{fig:dashboard}
  \end{minipage}
\end{figure}

Figure~\ref{fig:dashboard} shows a mock dashboard view for two datacenters (DC1 and DC2) in red state.

%\begin{figure}[h]
%  \vspace{-0.2cm}
%  \centering
%  \includegraphics[width=0.6\linewidth]{ansc_dashboard.png}
%  \vspace{-0.2cm}
%  \caption{ANSC operator dashboard mock.}
%  \vspace{-0.2cm}
%  \label{fig:dashboard}
%\end{figure}

\section{Discussion}
ANSC enables operators to prioritize fixes not by first-come incidents, but by systemic risk posture. For example, if one region shows multiple datacenters at elevated risk, capacity upgrade or preventative link replacement can be prioritized. Furthermore, ANSC integrates naturally with an AI-based recommendation system that advises SREs on mitigation: suggesting whether a failing device requires datacenter technician intervention, or whether automated incident response suffices. This shift allows planning teams to move from reactive firefighting toward measurable risk budgeting at both datacenter and regional scale.
To make progress visible, ANSC can be shown in two ways: overall posture (exposure ceiling) and its movement over time. Posture highlights structural risk, while movement reflects the operator’s effort.

\section{Related Work}
Past works such as network-telemetry~\cite{zhou2022evolvable}, Generative-RCA ~\cite{li2025coca}, diagnosis-RCA~\cite{li2024exchain} and B4~\cite{jain2013b4} have focused on monitoring and root-cause analysis. Systems like SCORE~\cite{agarwala2005score}, Sherlock~\cite{baset2007sherlock}, Net-assys~\cite{zolfaghari2014net} and network-anomaly-detection~\cite{bekerman2019knowing} leverage probabilistic models to reason under uncertainty, automatically diagnosing network issues by analyzing everything from component dependencies and path measurements to monitoring data confidence. Our contribution differs in focusing on \emph{proactive probabilistic scoring} rather than reactive diagnosis. SREbook~\cite{beyer2018site} provide foundational guidelines, but ANSC introduces a quantifiable score tailored to Clos capacity headroom. Unlike these systems, ANSC explicitly ties probability of future failures with current capacity margins, providing a forward-looking risk score.

\section{Conclusion and Future Work}
We have presented ANSC, a probabilistic scoring framework that forecasts systemic datacenter capacity risk. By combining capacity forecasting, historical failure rates, and normalized budgets, ANSC provides operators with clear prioritization signals. Future work includes deploying ANSC in production and evaluating operator response time reduction.

\appendix
\section{Ethics}
All experiments are conducted on infrastructure telemetry and simulated incidents; no human subjects or sensitive data are involved, and therefore no ethical issues arise.

\bibliographystyle{ACM-Reference-Format}
\bibliography{main}

\end{document}